# Tuneable spin injection in high-quality graphene with one-dimensional contacts

Victor H. Guarochico-Moreira,[1,2] Jose L. Sambricio,[1] Khalid Omari,[1] Christopher R. Anderson,[1] Denis A. Bandurin,[1] Jesus C. Toscano-Figueroa,[1,3] Noel Natera-Cordero,[1,3] Kenji Watanabe,[4] Takashi Taniguchi,[4] Irina V. Grigorieva,[1*] and Ivan J. Vera-Marun[1*]

[1]*Department of Physics and Astronomy, University of Manchester, Manchester M13 9PL, United Kingdom*

[2]*Escuela Superior Politécnica del Litoral, ESPOL, Facultad de Ciencias Naturales y Matemáticas, Campus Gustavo Galindo Km. 30.5 Vía Perimetral, P.O. Box 09-01-5863, Guayaquil, Ecuador*

[3]*Consejo Nacional de Ciencia y Tecnología (CONACyT), Av. Insurgentes Sur 1582, Col. Crédito Constructor, Alcaldía Benito Juarez, C.P. 03940, Ciudad de México, México*

[4]*National Institute for Materials Science, 1-1 Namiki, Tsukuba 305-0044, Japan*

**ABSTRACT**

Spintronics involves the development of low-dimensional electronic systems with potential use in quantum-based computation. In graphene, there has been significant progress in improving spin transport characteristics by encapsulation and reducing impurities, but the influence of standard two-dimensional (2D) tunnel contacts, via pinholes and doping of the graphene channel, remains difficult to eliminate. Here, we report the observation of spin injection and tuneable spin signal in fully-encapsulated graphene, enabled by van der Waals heterostructures with one-dimensional (1D) contacts. This architecture prevents significant doping from the contacts, enabling high-quality graphene channels, currently with mobilities up to 130,000 $cm^2V^{-1}s^{-1}$ and spin diffusion lengths approaching 20 µm. The nanoscale-wide 1D contacts allow spin injection both at room and at low temperature, with the latter exhibiting efficiency comparable with 2D tunnel contacts. At low temperature, the spin signals can be enhanced by as much as an order of magnitude by electrostatic gating, adding new functionality.

**Corresponding Authors**

*E-mail: irina.grigorieva@manchester.ac.uk

*E-mail: ivan.veramarun@manchester.ac.uk

Graphene is currently explored for potential applications on a variety of fields due to its exceptional physical properties[1] including high-quality electronic transport[2,3]. In spintronics, where the spin degree of freedom is used to store, transport and manipulate information, graphene has attracted interest as a spin transport channel[4,5]. There has been much progress in improving spin transport characteristics in graphene[6,7] but some challenges remain, such as the inhomogeneity in the potential profile within the channel[8,9], invasive tunnelling contacts[10,11] and impurities[12–14]. Therefore device architecture plays a key role to avoid the aforementioned challenges, for example via the realisation of high-quality spin channels[6] within van der Waals heterostructures[15]. Progress in this area opens an avenue towards exploiting ballistic conduction and spin coherence in graphene-based quantum spintronics[16–18].

Full encapsulation by hexagonal boron nitride (hBN) protects graphene from direct contact with lithographic polymers, whereas the self-cleaning process driven by van der Waals interactions limits any contamination present at the interfaces within small bubbles, ensuring atomically clean interfaces in the rest of the heterostructure[19]. For spintronics, we need to be able to inject spin information. This is traditionally achieved via magnetic tunnel contacts. The use of tunnel barriers has grown out of the need to overcome the so-called conductivity mismatch problem[20], which leads to a drastically reduced spin injection efficiency when the resistance of the contact is lower than the spin resistance of the channel[21], $R_s = \rho \lambda / W$, with $W$ the graphene channel width, $\lambda$ the spin relaxation length and $\rho$ the graphene sheet resistance. Nevertheless, standard 2D contacts, even with the use of a tunnel barrier, are known to introduce strong doping across the channel leading to inhomogeneity[22–24] and present challenges in the growth of barrier without pinholes, which lead to enhanced spin relaxation[5,25]. On the other hand, fully encapsulated graphene with 1D contacts[26] has been shown to produce exceptionally clean devices and localise the doping within ~ 100 nm near the contacts[26–28]. 1D contacts have recently enabled spin injection in graphene[29], albeit only at low temperature, with graphene channel mobilities below 30,000 cm$^2$V$^{-1}$s$^{-1}$ and contact polarisation with multiple inversions of polarity. Therefore, further study of this architecture is warranted.

Here, we report spin transport in high-quality graphene channels with low-temperature mobility up to ~ 130,000 cm$^2$V$^{-1}$s$^{-1}$, fully encapsulated by hBN layers, where a spin current is injected via nanoscale-wide 1D contacts. Spin precession measurements yield a quantitative understanding by extracting the channel spin relaxation length $\lambda$ and time $\tau_s$, and the 1D contacts' spin injection efficiency or polarisation $P$. The fabrication process produces homogeneous graphene channels, including 1D contacts that prevent substantial charge doping within the channel and offer a gate-tuneable contact resistance. While the spin polarisation is comparable to that from standard tunnel 2D contacts, spin transport can be electrostatically tuned into a mismatch-free spin injection regime. These elements lead to the realisation of a ballistic injection process via nanoscale-wide 1D contacts and spin transport in graphene with a spin relaxation length of ~ 18 µm and a long mean free path of ~ 1 µm, opening up possibilities for lateral spintronic elements that exploit quantum transport.

Our devices are heterostructures consisting of monolayer graphene encapsulated between two thin (< 20 nm) layers of hBN, with ferromagnetic contacts deposited directly onto narrow (~10 nm wide) strips of graphene at the sides of the channel (see Figs. 1(a) and (b)). Details of device fabrication and characterisation are given in Supporting Information Section 1. Briefly, we use the dry-peel transfer technique[30] to prepare van der Waals heterotructures on a Si/SiO$_2$ substrate. We then use standard electron beam lithography to pattern a hard polymer mask which defines the channel geometry by using reactive ion etching[31]. Due to different etch rates for hBN and graphene, a ~ 10 nm wide step can be seen in the profile of a hBN-graphene-hBN edge (green line in Fig. 1(c)). This step corresponds to a narrow graphene ledge[28] where the 1D contact is formed. For comparison a profile of an hBN-hBN edge is shown (red line in Fig. 1(c)) where this step is not visible. Finally, we deposit ferromagnetic contacts that pass over the channel, creating electrical connections to only the edges of the graphene layer (see inset of Fig. 1(a)).

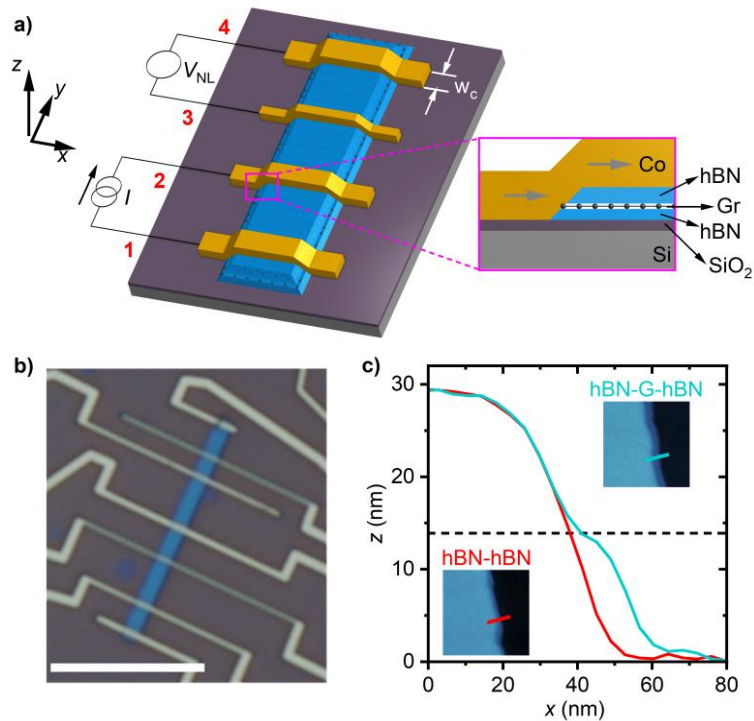

**FIG. 1. Device fabrication and characterisation. a**, 3D schematic representation of an hBN-graphene-hBN channel with magnetic 1D contacts connected in a 4-probe nonlocal measurement configuration. The inset shows a cross sectional view. **b**, Optical microscopy image of a typical device. Scale bar 10 µm. **c**, Height profiles of a channel's edge. Red (green) line shows the hBN-hBN (hBN-graphene-hBN) profile from the atomic force microscopy (AFM) image at the bottom-left (top-right) inset. Size of the AFM scan window is 500 nm x 500 nm. The horizontal black dotted line indicates the position where the graphene lies between the top and the bottom hBN.

Nine devices, labelled **A-I**, were studied using charge and spin transport measurements. All devices have shown qualitatively similar behaviour, both at room and low temperature (20 K). In order to characterise charge transport we measured the four-probe resistance of graphene as a function of its charge carrier density, $n$, by using a back-gate voltage applied between the highly doped Si substrate and the graphene channel (see inset in Fig. 1(a)). The curves in Fig. 2(b) show the conductivity $\sigma = 1/\rho$ of graphene at low and room temperature for device **A**. Our devices show a uniform level of doping, within $\pm 7 \times 10^{11}$ cm$^{-2}$. Given the lack of a defined doping polarity it was not possible to attribute any particular doping at the graphene edges originating from the fabrication process. To evaluate the electronic quality we extracted the corresponding field-effect mobility of the graphene channel as $\mu_{FE} = (d\sigma/dn)/e$, as shown in Fig. 2(b), at moderate carrier densities $|n| \sim 1\times10^{12}$ cm$^{-2}$. For device **A** the mobility is then 45,000 cm$^2$V$^{-1}$s$^{-1}$ at room temperature and 79,000 cm$^2$V$^{-1}$s$^{-1}$ at 20 K. The mobilities of our devices, typically within the range of 20,000 to 130,000 cm$^2$V$^{-1}$s$^{-1}$ (see Supporting Information Section 2), are significantly higher than previous graphene-based spintronic devices[9,10,12,32] that exhibited mobilities < 20,000 cm$^2$V$^{-1}$s$^{-1}$, the majority of which have used only a partial hBN encapsulation, whereas we ensure this full hBN encapsulation throughout the spin transport channel. The charge diffusion coefficient (and corresponding mean free path) for representative devices is obtained from the graphene sheet resistance via the Einstein relation, $D = 1/(\rho e^2 \nu)$, with $\nu$ the density of states for single layer graphene (see Fig. 2(d)).

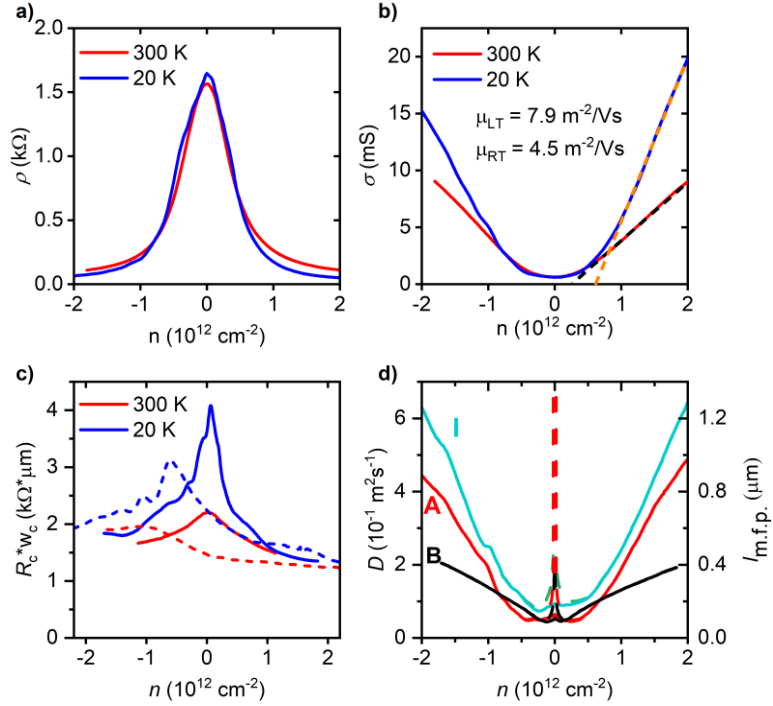

**FIG. 2. Charge transport in devices with 1D contacts. a**, **b**, Graphene sheet resistance (**a**) and conductivity (**b**) vs. carrier density. Panel **b** shows the extracted field-effect mobility. Data in panels **a**, **b** are for device **A**. **c**, Contact resistance-length product as a function of carrier density for two 1D contacts (continuous and dashed lines). In panels **a-c** blue and red curves are for 20 K and 300 K, respectively. **d**, Diffusion coefficient and mean free path as a function of carrier density, at 20 K. Data in panel **d** correspond to three representative devices: **I** (cyan), **A** (red) and **B** (black).

To characterise the 1D contact resistance $R_c$ as a function of carrier density $n$, as shown in Fig. 2(c), we use a three-probe geometry (see Supporting Information Section 1). The contacts exhibit typical $R_c$ values within the range of $3 - 15$ k$\Omega$, which implies they should be able to inject efficiently spins into a graphene channel with $R_s < 10$ k$\Omega$. Some contacts had electrical access via their two ends, see Fig. 1(b), which allowed a four-probe measurement to quantify the contribution of the lead series resistance ($\sim 200$ $\Omega$, Supporting Information Section 3). The magnetic 1D contacts present charge transport characteristics consistent with those of reported non-magnetic ones[26] (see Supporting Information Section 4). Among these are: (i) an inverse scaling of $R_c$ with the width of the contact $w_c$ (see Fig. S4(a)), (ii) a negligible temperature dependence of $R_c$ for high carrier density (see Fig. S4(c)), and (iii) a sizeable dependence on carrier density, with a moderate electron-hole asymmetry and a maximum $R_c$ near the Dirac point (see Fig. 2(c) and Fig. S4(b)). The electron-hole asymmetry of the contact in Fig. 2(c), showing a somewhat larger resistance for transport in the hole regime and a maximum $R_c$ at $n \lesssim 0$, indicates the presence of an n-doped region adjacent to the metal electrode, consistent with the difference between the work functions of the metal (Co) and graphene[33].

Spin transport is characterised by spin-valve and spin precession (Hanle) measurements in a standard non-local geometry, as shown in Fig. 1(a). We inject a spin-polarised current $I$ into graphene through contacts 1 and 2, and measure a non-local voltage $V_{NL}$ between contacts 3 and 4. The non-local resistance is defined as $R_{NL} = V_{NL}/I$. The spin valve signal is given by the difference between the two distinct levels corresponding to the parallel and antiparallel alignment of the injector and detector electrodes, $\Delta R_{NL} = R_{NL}^{P} - R_{NL}^{AP}$. Spin signals were measured for different separations between injector and detector, $L$, ranging from 2 to 15 µm (see Supporting Information Section 7). An increase of approximately one order of magnitude in $\Delta R_{NL}$ from room to low temperature was observed, as shown in Figs. 3(a) and (b), with the latter reaching $\Delta R_{NL} > 1$ Ω. This strong temperature dependence is distinct from the weaker dependence observed in standard tunnel 2D contacts[5], indicating a different transport mechanism for spin injection in 1D contacts.

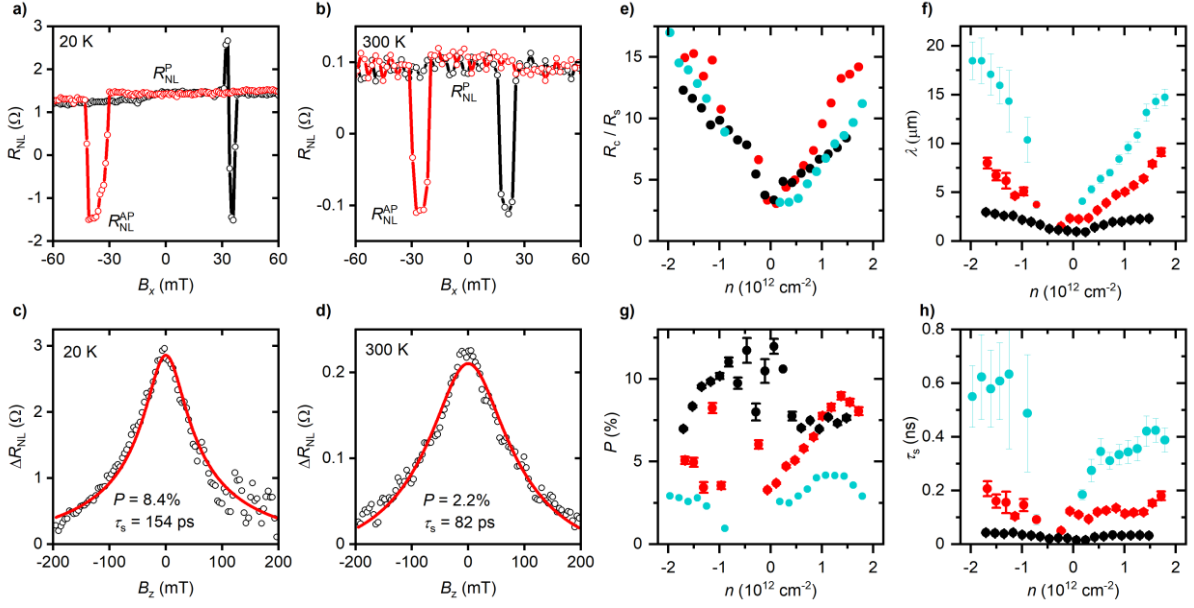

**FIG. 3. Spin transport in devices with 1D contacts. a**, **b**, Spin valve measurements. Black (red) curve represents the up (down) sweep of in-plane magnetic field. **c**, **d**, Hanle spin precession measurements. Red curve is a fit to the Bloch equation, using the parameters shown in the panel. Data in panels **a-d** are for device **A**, with $L = 2.4$ µm. Panels **a**, **c** (**b**, **d**) are for low (room) temperature. **e,f,g,h**, Spin transport parameters. Contact resistance to channel spin resistance ratio (**e**), spin relaxation length (**f**), spin polarisation (**g**), and spin relaxation time (**h**) as a function of carrier density, at 20 K. Data in panels **e-h** correspond to the same three representative devices as in Fig. 2d: **I** (cyan), **A** (red) and **B** (black).

The observation of a Hanle signal[34] enables us to confirm the presence of spin transport and rule out spurious contributions which hindered previous efforts in devices with 1D contacts[32]. Here $R_{NL}^P$ and $R_{NL}^{AP}$ are measured while sweeping an external magnetic field applied perpendicular to the device plane ($B_z$). This causes the injected spins, having a polarisation along the x-direction (see Fig. 1(a)), to precess within the x-y plane while moving within the channel. The resulting $\Delta R_{NL}(B_z)$ (see Figs. 3(c) and (d)) is analyzed using a solution to the steady-state Bloch equation[34,35],

$$\frac{d\vec{\mu}_s}{dt} = D\nabla^2\vec{\mu}_s - \frac{\vec{\mu}_s}{\tau_s} + \vec{\omega}_L \times \vec{\mu}_s = 0, \qquad (1)$$

where $\vec{\mu}_s$ is the spin accumulation within the channel and $D$ is the diffusion coefficient, the latter being equal to the charge diffusion coefficient in the absence of spin Coulomb drag[36]. The term $\vec{\omega}_L \times \vec{\mu}_s$ describes spin precession under an external magnetic field $\vec{B}$, where the Larmor frequency is given by $\vec{\omega}_L = g\mu_B/\hbar \vec{B}$, with the Lande factor g = 2 and $\mu_B$ the Bohr magneton. Therefore the spin transport parameters ($\tau_s$, $D$ and $\lambda = \sqrt{D\tau_s}$) of the channel are extracted.

The spin signal $\Delta R_{NL}$ depends on the spin polarisation of the magnetic 1D contacts, $P$. In the absence of spin precession ($B_z = 0$) the spin signal is described by a balance of spin injection and relaxation within the channel, given by[34],

$$\Delta R_{\text{NL}} = \frac{P^2 \rho \lambda}{W} e^{-\frac{L}{\lambda}}, \qquad (2)$$

with $L$ the injector-detector distance and $W$ the channel width. The channel parameters for charge ($\sigma \propto D$) and spin ($\tau_s$) transport have a moderate twofold increase at low temperature, see Figs. 2 and 3. A significant increase in conductivity at low temperature, in the high-density regime, is expected in high-quality graphene due to the absence of acoustic-phonon scattering[26]. Furthermore, $P$ has a fourfold increase at low temperature, as confirmed by the parameters extracted from Hanle measurements. These observations indicate that the strong temperature dependence seen in $\Delta R_{\text{NL}}$ is dominated by the polarisation of the contacts, via the scaling $\Delta R_{\text{NL}} \propto P^2$.

The identification of an n-doped graphene region next to our 1D contacts from the electron-hole asymmetric response of $R_c$, together with the observation of a more electron-hole symmetric channel resistance, as seen in Fig. 2(a), can be understood by considering the distinct geometries and length scales involved in both measurements. For the case of the contact resistance, the direct metal-graphene contact occurs only within the 1D edge, identified as the < 10 nm step in the heterostructure profile, see Fig. 1(c). Furthermore, the n-doped graphene region in direct contact with the metal is expected to extend to just ~ 100 nm[37,38] to the rest of the channel. This doped region only exists near the 1D contact, which in our devices has a nominal width $w_c$ of 100—350 nm. The n-doped region being localised in the vicinity of the 1D edge is in stark contrast to standard 2D junctions which cover the full width of the channel and lead to substantial inhomogeneity[22–24]. Here, the geometry of the 1D contacts plays a key role. On the other hand, for the case of the channel sheet resistance, most of the channel is undoped, except for the nanoscale region in direct vicinity of the 1D edge. Therefore the channel resistance, probed at a length scale $L, W \geq 1$ μm, exhibits a more electron-hole symmetric response.

A further characteristic that distinguishes these magnetic 1D contacts with nanoscale geometry from the behaviour in standard 2D junctions[33,38], and which is hitherto unaddressed in non-magnetic[26,28] or magnetic[29,32] 1D contacts, is a non-metallic increase in resistance at low temperature. All measured 1D contacts (see Fig. 2(c)) consistently exhibit a marked carrier density-dependent contact resistance at low temperature, indicating that $R_c$ varies with the Fermi energy. This dependence is considerably weaker at room temperature. At low carrier density, $R_c$ typically exhibits up to a twofold increase at low temperature. This behaviour is distinct from that of both the graphene channel sheet resistance which is essentially temperature-independent (see Fig. 2(a)), and from typical 2D metal-graphene ohmic junctions[33,38]. Only at high carrier density is this temperature dependence of $R_c$ reduced, consistent with the weak temperature dependence observed in 1D contacts with[26] $w_c > 1$ μm.

The description of contact resistance in 2D metal-graphene junctions involves two processes: carrier transport from the metal to the (doped) graphene region in direct contact with the metal, and transport from the doped graphene region to the rest of the (undoped) graphene channel. The transmission for those two processes gives rise to the observed contact resistance[38]. The first process is constrained by the number of conduction modes in the n-doped graphene region, which is limited by the contact width. Within this description, our nanoscale contacts limit the number of conduction modes in the junction and reduce the conductance at the metal-graphene interface, so that ballistic transport across a width of ~ 100 nm at low temperature would still lead to a sizeable $R_c$[28,38]. The latter enables spin injection, which in graphene has traditionally required the use of tunnel barriers[20] in order to overcome the conductivity mismatch problem[21] that arises when the injected spins are backscattered into the magnetic contact where they rapidly relax their spin orientation. In this case, the linear room temperature response corresponds to the thermally smeared Sharvin resistance[39]. With regards to the second process, transport occurs across a potential profile in graphene of ~ 100 nm length scale, smaller than the mean free path in most of our channels. Crucially, this energy-dependent process is tuneable in 1D junctions since, unlike 2D junctions where the Fermi level of graphene under the contacts is strongly pinned, the Fermi level of graphene near a 1D junction can be tuned efficiently[29], resulting in a tuneable contact resistance. This difference between 2D junctions and 1D junctions derives from their distinct dimensionality and scale of the metal-graphene interface (< 10 nm) and the doped graphene region (~ 100 nm), see Fig. 4(b).

Figs. 3(f), (g) and (h) show spin transport parameters extracted from Hanle measurements, for three representative devices at 20 K. Device **I** (cyan data) has a high electronic quality close to the upper limit of our mobility range, with $\mu_{FE} = 130{,}000$ cm$^2$V$^{-1}$s$^{-1}$. Device **A** (red data) also has a high electronic quality, with $\mu_{FE} = 79{,}000$ cm$^2$V$^{-1}$s$^{-1}$, whereas device **B** (black data) has a lower quality ($\mu_{FE} = 27{,}000$ cm$^2$V$^{-1}$s$^{-1}$). As shown in Fig. 3(g), we achieve contact polarisation values of up to ~ 10 %, comparable with standard tunnel contacts on graphene[25,34] (see also Fig. S5(c)).

The spin relaxation time $\tau_s$ and length $\lambda$, shown in Figs. 3(f) and (h), describe the spin transport within the graphene channel. Both parameters exhibit an approximate electron-hole symmetry, consistent with the graphene sheet resistance (see Fig. 2(a)). Moreover, their magnitude correlates with the electronic quality, with $\tau_s$ and $\lambda$ for device **I** being about twice as large as those for device **A**, while the parameters for device **A** are ca. three times as great as for device **B**. This observation indicates spin relaxation dominated by the Elliot-Yafet mechanism[40,41], where $\tau_s$ increases with the diffusion coefficient $D$ (see Fig. 2(d)). A priori, we cannot rule out contributions from other mechanisms, as previous studies of graphene-hBN heterostructures have found contributions from both Elliot-Yafet and D'Yakonov-Perel mechanisms[42], and there might also be other sources of spin relaxation due to the direct ferromagnetic contacts to graphene[43]. To ascertain that spin absorption at the ferromagnet-graphene interfaces did not play a significant role, we used the extended expression for $\Delta R_{NL}$ that included contact resistances[35]. The corresponding fit is shown in Fig. S7 (Supplementary Information). It is clear that, with our experimental accuracy, it is indistinguishable from the fit obtained using eq. (1), and that the extracted values of $\tau_s$ are not significantly affected. Further indication for the role of the Elliot-Yafet mechanism is obtained by studying the carrier density dependence. For all our devices, $\tau_s$ and $\lambda$ increase with increasing carrier concentration, having a minimum close to the Dirac point. This behaviour, also observed in single layer graphene using 2D contacts, has been attributed to the Elliot-Yafet mechanism[44,45] and results in a linear scaling between spin relaxation time $\tau_s$ and momentum relaxation time $\tau_p$ (see Supporting Information Section

6). We have observed spin transport across distances of 15 µm, while crossing several (non-invasive) 1D contacts, only limited by device dimensions (see Fig. S7). Overall, the demonstration of spin transport parameters reaching values up to $D \sim 0.7$ m$^2$s$^{-1}$ and $\lambda \sim 18$ µm open up the exploration of lateral spintronic architectures involving long distance spin transport[42], now within high-quality and homogeneous channels.

We note that, despite the clear indications that Elliot-Yafet mechanism of spin relaxation plays a dominant role in our devices, the observed $\tau_s$ is rather short, in fact shorter than in many devices with 2D contacts and of lower quality reported in literature[10,12]. In the latter case $\tau_s \sim 2$ ns or more have been observed, while the maximum spin lifetime in our device **I** with the highest mobility is only ~0.7 ns. We speculate that the reason is likely to be due to the specific geometry of spin injection through 1D contacts: here spin-polarized electrons enter from the opposite sides of a graphene channel which is followed by propagation in the perpendicular direction along the channel. In this case spin transport becomes essentially 2D and may not be accurately described by the standard 1D Hanle model (where spins are injected uniformly across the channel and continue propagation in the same direction).

We evaluate the degree of conductivity mismatch via the ratio of contact resistance, $R_c$, to spin resistance of the graphene channel, $R_s$, where a ratio $R_c/R_s \gg 1$ indicates a regime of efficient spin injection[5,11]. In graphene spintronic devices, the ratio $R_c/R_s$ has been tuned by using an electrostatic back-gate via the carrier density dependence of $\rho$ and $\lambda$, whereas for tunnel junctions $R_c$ typically exhibits a weak dependence[5,25]. In our work, not only the channel resistance, but also the 1D contacts exhibit a dependence on carrier density (see Fig. 2(c)) which, as we show below, can be used to overcome the impedance mismatch problem. As shown in Fig. 3(e), the ratio remains $R_c/R_s > 1$ for all carrier densities, increases with carrier density, and is tuneable to values $R_c/R_s \gg 1$ at high density.

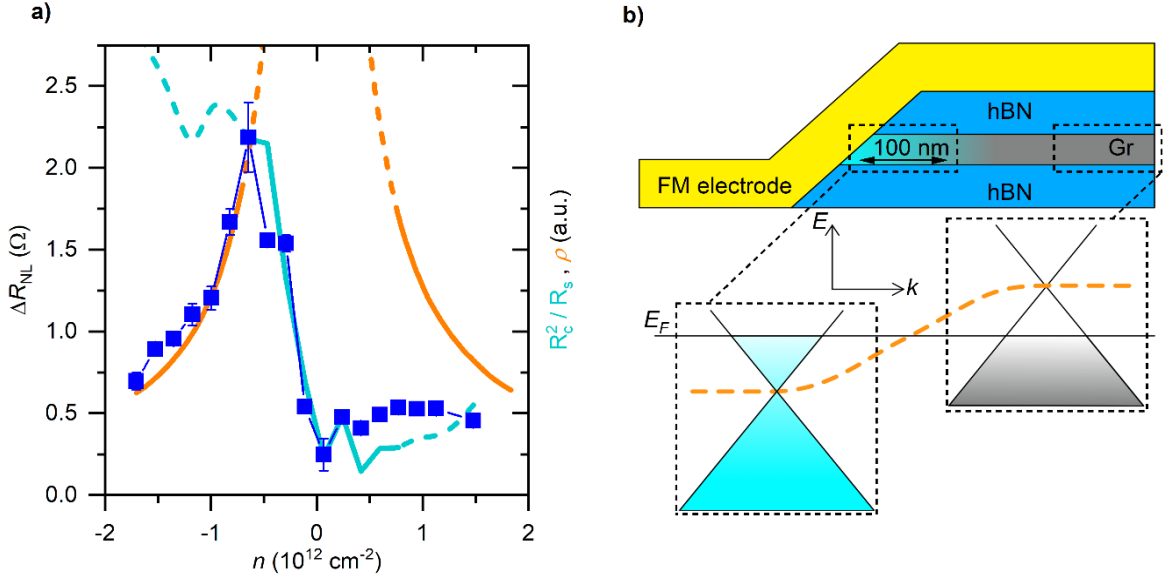

**FIG. 4. Tuneable spin injection efficiency at 20 K. a**, Spin valve signal as a function of carrier density for device **C** (blue squares) with $L = 5.6$ µm. The orange (cyan) line represents the graphene sheet resistance (contact to spin resistance ratio), with each line turning solid to indicate a similar scaling as the spin valve signal. **b**, Schematic representation of a magnetic 1D contact to an hBN-graphene-hBN channel, with the orange dashed line depicting the position of the neutrality point within graphene.

To further evaluate the spin injection efficiency and its tuneability, we discuss the spin valve signal, $\Delta R_{NL}$. As shown in Fig. 4(a), $\Delta R_{NL}$ has a minimum at the Dirac point, consistent with the minimum observed for the spin transport parameters (Figs. 3(f) & (h)). Away from the Dirac point, $\Delta R_{NL}$ exhibits a non-monotonic behaviour: in the hole regime, the signal increases and reaches a maximum for $n \sim -0.8 \times 10^{12}$ cm$^{-2}$, after which it decreases. For the electron regime, this behaviour is much less pronounced and the spin signal remains comparatively low. This overall behaviour is understood by considering two distinct regimes of conductivity matching[25]. First, at high carrier density, where $R_c/R_s \sim 10$, there is a regime free of conductivity mismatch. Here, $\Delta R_{NL}$ is insensitive to the value of $R_c$ and it scales as $\Delta R_{NL} \propto \rho$ (see Eq. 2). This is observed in the hole regime, as indicated by the orange curve in Fig. 4(a). On the other hand, at low carrier density $R_c/R_s \sim 3$, see Fig. 3(e). The latter corresponds to an intermediate regime where, although nominally in a conductivity-matched condition, the spin transport is still sensitive to the value of $R_c$. In this case $\Delta R_{NL}$ exhibits a scaling similar to that in the mismatch regime[46] $\propto R_c^2/R_s$. As indicated by the cyan line in Fig. 4(a), the latter regime accounts for the reduction in $\Delta R_{NL}$[5,11] at low carrier density and in the electron regime.

This device architecture demonstrates efficient spin injection in graphene using nanoscale-wide 1D contacts, reproducibly across several devices. The ballistic spin injection process allows for observation of the Hanle effect[32], tuneable spin signal and achieves a mismatch-free regime at moderate carrier density. The spin signal shows a consistent behaviour across all devices, with a minimum near the Dirac point. This observation is compatible with the proposal of magnetic proximity at the Co-graphene interface[29,33], where a reversal in the polarity of the spin polarisation is expected near the Dirac point, implying no spin polarisation and thus no spin signal near the Dirac point. The fact that our devices do not show any inversion of the spin signal as a function of carrier density is consistent with a homogeneous potential profile within the graphene channel, where both injector and detector contacts would reverse their polarisations at similar carrier density. The demonstration of spin transport in graphene with a large mean free path of ~ 1 µm at low temperature, comparable to device dimensions, while ensuring sizable spin relaxation lengths, is a key advance in the development of low-dimensional spintronic systems approaching[47] the high electronic mobility of state of the art charge-based devices[26,48].

**Supporting Information.** The Supporting Information contains details on the overall device fabrication and characterisation, characterisation of additional devices, contributions to the contact resistance, behaviour of the one-dimensional magnetic contacts, an additional device with analysis of Hanle spin precession, exploring the mechanism of spin relaxation, and long distance spin transport (PDF). This material is available free of charge via the Internet at http://pubs.acs.org.

## ACKNOWLEDGMENTS

We thank Dr. Moshe Ben Shalom for useful discussions. We acknowledge support from the European Union's Horizon 2020 research and innovation program under Grant Agreement Nos. 696656 and 785219 (Graphene Flagship Core 2), and from the Engineering and Physical Sciences Research Council (UK) EPSRC CDT Graphene NOWNANO EP/L01548X. V.H.G.-M. acknowledges support from the Secretaría Nacional de Educación Superior, Ciencia y Tecnología (SENESCYT) for the PhD scholarship under the program "Universidades de Excelencia 2014" (Ecuador). J.L.S. and D.A.B. acknowledge support from the FP7 Marie Curie Initial Training Network "Spintronics in Graphene" (SPINOGRAPH). K.O. and I.J.V.M. acknowledges support from the FP7 FET-Open Grant 618083 (CNTQC). C.R.A. acknowledges support from the EPSRC Doctoral Training Partnership (DTP). J.C.T.-F. and N.N.-C. acknowledge support from the Consejo Nacional de Ciencia y Tecnología (México). Research data are available from the authors upon request.

# SUPPORTING INFORMATION

# Tuneable spin injection in high-quality graphene with one-dimensional contacts

Victor H. Guarochico-Moreira,[1,2] Jose L. Sambricio,[1] Khalid Omari,[1] Christopher R. Anderson,[1] Denis A. Bandurin,[1] Jesus C. Toscano-Figueroa,[1,3] Noel Natera-Cordero,[1,3] Kenji Watanabe,[4] Takashi Taniguchi,[4] Irina V. Grigorieva,[1*] and Ivan J. Vera-Marun[1*]

[1]*Department of Physics and Astronomy, University of Manchester, Manchester M13 9PL, United Kingdom*

[2]*Escuela Superior Politécnica del Litoral, ESPOL, Facultad de Ciencias Naturales y Matemáticas, Campus Gustavo Galindo Km. 30.5 Vía Perimetral, P.O. Box 09-01-5863, Guayaquil, Ecuador*

[3]*Consejo Nacional de Ciencia y Tecnología (CONACyT), Av. Insurgentes Sur 1582, Col. Crédito Constructor, Alcaldía Benito Juarez, C.P. 03940, Ciudad de México, México*

[4]*National Institute for Materials Science, 1-1 Namiki, Tsukuba 305-0044, Japan*

**Corresponding Authors**

*E-mail: irina.grigorieva@manchester.ac.uk

*E-mail: ivan.veramarun@manchester.ac.uk

## Section 1. Device fabrication and characterisation

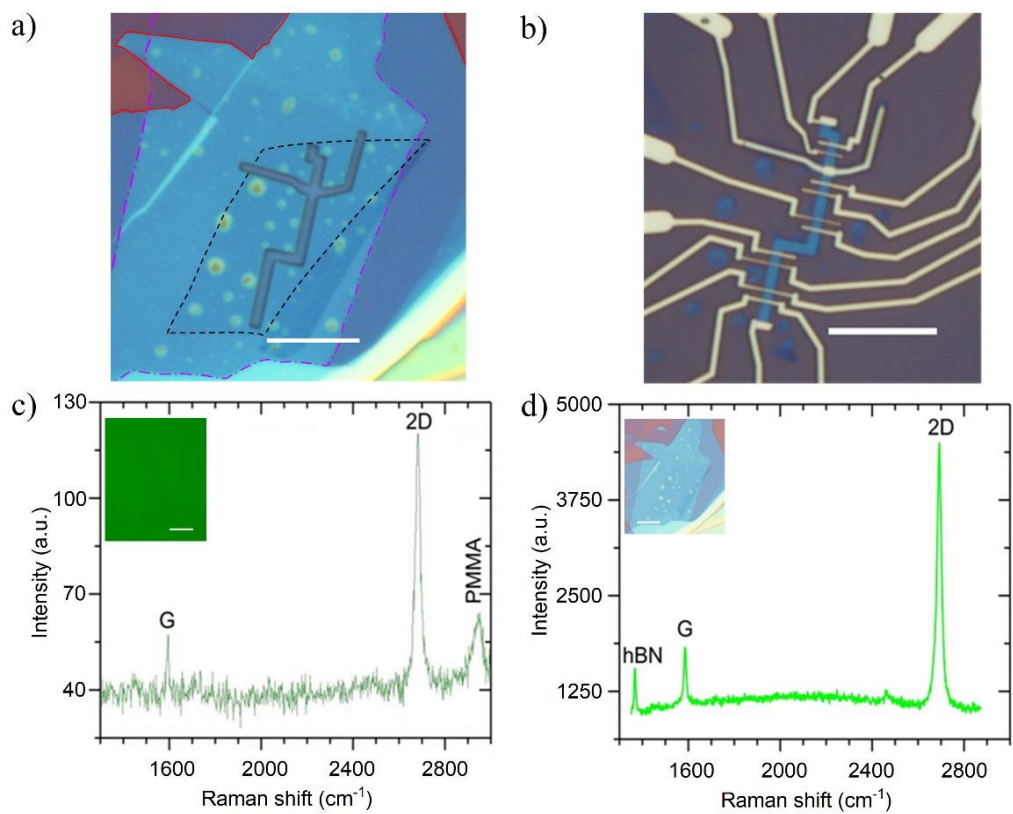

**Figure S1.- Device fabrication and Raman characterisation. a,** Optical microscopy picture of our hBN-graphene-hBN stack with a PMMA mask on top to define the encapsulated graphene channel. Solid red, dot-dashed purple and dashed black lines represent the boundaries of the top hBN, bottom hBN and graphene, respectively. **b,** Optical microscopy picture of one typical device, after depositing the magnetic contacts. **c,** Raman spectrum of graphene on a PMMA/PMGI stack. Inset shows a green-filter optical image of the graphene flake in the stack in (**a**). **d,** Raman spectrum of the hBN-graphene-hBN heterostructure. Inset shows an optical image of the stack in (**a**). All scale bars are 10 µm.

The hBN-graphene-hBN heterostructures were assembled on a highly p-doped Si substrate, with a $SiO_2$ layer of 290 nm thickness. Figure S1a depicts an optical image of a typical hBN-graphene-hBN stack assembled using the dry-peel transfer technique[1]. Graphene and hBN were mechanically exfoliated on top of a polymer stack consisting of poly(methylglutarimide) (PMGI) and poly(methyl methacrylate) (PMMA), previously coated onto a Si substrate. The sacrificial PMGI layer is dissolved with a water-based solvent (MF319) while the top layer together with the desired hBN flake is lifted off from the Si substrate. The resulting PMMA membrane is placed onto a metal ring and loaded face down into a setup with micromanipulators to align the top hBN with the graphene flake, previously prepared on the polymer stack. The graphene flake is picked up by the top hBN flake attached to the PMMA membrane, via the dry-peel transfer technique[1]. This structure is aligned with a second (bottom) hBN flake, previously exfoliated on a Si substrate, to finalise the assembly of the heterostructure. The PMMA membrane is dissolved in acetone, followed by annealing at 300°C in an atmosphere of $Ar/H_2$ gas mixture.

A self-cleaning process[2] coerces any contamination (hydrocarbons and absorbed water), present at the surfaces of graphene and hBN, to cluster into submicron-sized bubbles during assembly. This ensures atomically clean interfaces in the majority of the heterostructure. The bubbles are identified by optical and atomic force microscopy and thus can be avoided when defining the channel's geometry (figure S1a). Electron beam lithography (EBL) is used to pattern a PMMA hard mask to define the channel, which is created using reactive ion etching with a mixture of $CHF_3$ and $O_2$ gases[3]. Next, we used a second step of EBL to pattern the contacts in a PMMA resist, followed by deposition of Co by electron-beam evaporation under a base pressure vacuum of $10^{-6}$ mbar, as shown in figure S1b.

The characterisation of our devices begins with confirmation that a single layer of graphene has been isolated. This is done using Raman spectroscopy, either following the exfoliation of graphite on PMMA (figure S1c) or following the assembly of the stack (figure S1d). Both spectra show the G and 2D peaks characteristic of single layer graphene (at ~1580 and ~2700 $cm^{-1}$ respectively). Additionally we observe peaks characteristic of PMMA or hBN (at ~1340 and ~2950 $cm^{-1}$ respectively).

For electrical characterisation, all devices were measured in a cryostat, under a vacuum atmosphere of pressure $< 6 \times 10^{-7}$ mbar, using standard lock-in techniques at low-frequency ($<$ 20 Hz). To electrostatically gate the graphene and to bias the contacts we used dc sources. The gate voltage, $V_{bg}$, induces a charge carrier density $n = C_g(V_{bg} - V_D)/e$, where $e$ is the elementary charge, $C_g$ is the geometrical gate capacitance per area, and $V_D$ is the gate voltage for the maximum of $\rho$. We determine $C_g$ by considering two dielectrics in series, the $SiO_2$ and the bottom hBN, the latter with a thickness determined by AFM. We extract the field-effect mobility of the graphene channel as $\mu_{FE} = (d\sigma/dn)/e$, with $\sigma = 1/\rho$. The diffusion coefficient is extracted from the Einstein relation, $D = 1/(\rho e^2 \nu)$, by using the density of states for single layer graphene, $\nu = \sqrt{g_s g_v n}/\sqrt{\pi}\hbar v_F$, where $g_s$ and $g_v$ are the spin and valley degeneracy respectively, and $v_F = 10^6$ m/s the Fermi velocity. The mean free path is given by $l_{m.f.p.} = 2D/v_F$, as shown in Fig. 2d.

Following the notation in Fig. 1a, to measure the $R_c$ of contact 2 in a three-probe geometry, current is applied via contacts 2 and 1 while voltage is measured between contacts 2 and 3. With this geometry we measure contributions of the 1D contact junction resistance, the spread resistance within the graphene channel, and the series lead resistance. The first contribution dominates $R_c$ and is discussed in the main text. Both the second contribution, due to the detection of charge current (Maxwell) spreading resistance[4,5], and the third contribution, due to the metallic electrode and measurement setup lead, are found to be negligible (see Supporting Information Section 3). For the evaluation of conductivity mismatch we use the $R_c$ values of both the injector and detector electrodes[6].

For the non-local geometry, the electrical connections are schematically shown in Fig. 1a, as described in the main text. Contacts 1 and 4 are generally chosen to be at the ends of the channel, so that spin transport is dominated only by contacts 2 (injector) and 3 (detector). A magnetic field is applied along the length of the magnetic electrodes (in-plane $B_x$) to configure the injector and detector electrodes in a parallel or an antiparallel state, corresponding to two distinct levels of the non-local resistance, $R_{NL}^{P}$ and $R_{NL}^{AP}$ (see Fig. 3a and Fig. 3b). This geometry separates the spin current from the (drift) charge current, reducing spurious magnetoresistance effects.

**Section 2. Characterisation of additional devices**

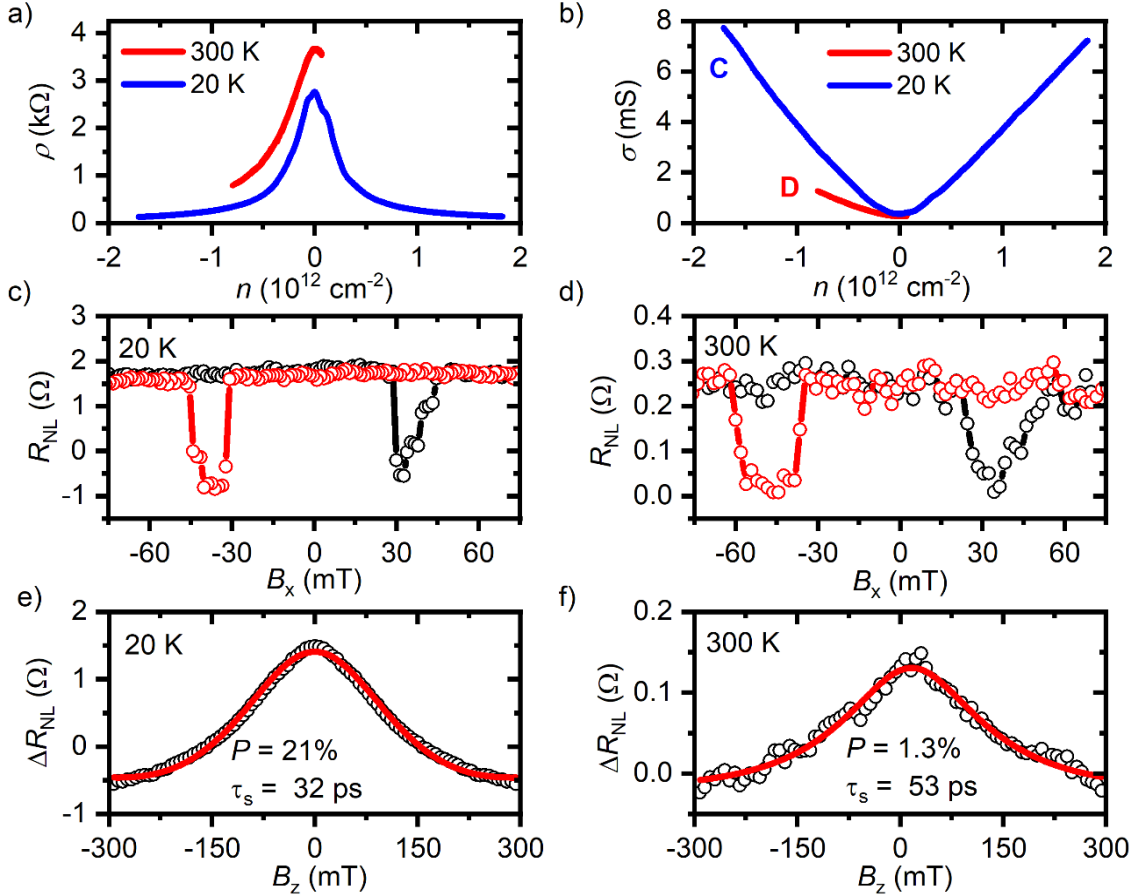

**Figure S2. Electrical Characterisation for device C (at 20 K) and D (at 300 K). a,** Dependency of graphene resistivity on charge carrier density. **b,** Dependency of conductivity on charge carrier density. **c, d,** Spin valve measurements at $n \sim -0.5 \times 10^{12}$ cm$^{-2}$. The black (red) curve is for the up (down) sweep of the in-plane magnetic field. **e, f,** Hanle spin precession curves at $n \sim -0.5 \times 10^{12}$ cm$^{-2}$. The solid lines are fits to the one-dimensional Bloch equation.

In this section we present data from the electrical characterisation of additional devices. Figure S2 shows the charge and spin transport of device C at 20 K and device D at 300 K. Device C has a distance between injector and detector of 5.6 µm and device D a distance of 1.5 µm. We typically observe the magnitude of the spin signal to be in the range of few hundreds of milliohms at 300 K, whereas at 20 K these values are one order of magnitude greater. This scaling appears to be dominated by the polarization of the magnetic contacts rather than by the spin relaxation in the channel (discussed in the main text and later here in Section 4).

A further summary of device geometry, layer thicknesses, and charge transport parameters for all devices is presented in table S1. As we can see, the mobility in our devices spans a range from ~ 10,000 cm$^2$V$^{-1}$s$^{-1}$ to ~ 50,000 cm$^2$V$^{-1}$s$^{-1}$ at 300 K. Whereas at 20 K this range spans from ~ 20,000 cm$^2$V$^{-1}$s$^{-1}$ to ~ 130,000 cm$^2$V$^{-1}$s$^{-1}$ (see table S1). We attribute these variations to the quality of edges, small bubbles and effects of strain. The last two have been greatly avoided thanks to the progressive improvement of the stacking process[7].

| Device | b-hBN (nm) | t-hBN (nm) | $L$ (µm) | $W$ (µm) | µ$_{FE}$ RT (cm$^{-2}$V$^{-1}$s$^{-1}$) | µ$_{FE}$ LT (cm$^{-2}$V$^{-1}$s$^{-1}$) | $n^*$ (cm$^{-2}$) | $D$ (m$^2$s$^{-1}$) | $l_{m.f.p.}$ (µm) |
|---|---|---|---|---|---|---|---|---|---|
| A | 6 | 9 | 2.4 | 1.1 | 45,000 | 79,000 | 3.5x10$^{11}$ | 0.47 | 0.93 |
| B | 12 | 15 | 5.1 | 1.2 | 19,000 | 27,000 | 1.5x10$^{11}$ | 0.20 | 0.40 |
| C | 12 | 15 | 5.6 | 1.2 | 19,000 | 30,000 | 1.4x10$^{11}$ | 0.20 | 0.41 |
| D | 10 | 12 | 1.5 | 1 | 16,000 | 17,000 | 3x10$^{10}$ | 0.06 | 0.12 |
| E | 5 | 15 | 2.6 | 1.2 | 12,000 | 14,000 | 4x10$^{10}$ | 0.06 | 0.13 |
| F | 7 | 14 | 1 | 1 | 12,000 | 16,000 | 1.3x10$^{11}$ | 0.07 | 0.14 |
| G | 12 | 15 | 12.7 | 1.2 | 18,000 | 28,000 | 1.6x10$^{11}$ | 0.20 | 0.39 |
| H | 6 | 9 | 3.1 | 2.4 | 41,000 | 85,000 | 3.3x10$^{11}$ | 0.52 | 1.05 |
| I | 6 | 9 | 3.1 | 2.4 | 47,000 | 130,000 | 3.3x10$^{11}$ | 0.64 | 1.3 |

**Table S1.- Summary of device parameters**. The field-effect mobility, µ$_{FE}$, is extracted for most devices at $|n|$ ~ 1—1.5x10$^{12}$ cm$^{-2}$, with the exception of devices **D** and **E** where this is done at $n$ = -0.5x10$^{12}$ cm$^{-2}$. The residual carrier density, $n^*$, charge diffusion coefficient, $D$, and the mean free path, $l_{m.f.p.}$, are extracted at 20 K. For most devices $D$ and $l_{m.f.p}$ are extracted at $|n|$ = 2x10$^{12}$ cm$^{-2}$, their maximum values, except for **D** and **E** where these parameters are extracted at $n$ = -0.5x10$^{12}$ cm$^{-2}$.

All the devices included in Table S1 demonstrated spin transport. Their characteristics were generally in line with our results discussed in the main text. In particular, they exhibited a strong temperature dependence, with spin-valve signal on the order of ohms at 20 K and a decrease by about one order of magnitude at room temperature. Table S2 presents a summary of the non-local spin valve signals observed both at 20 K and at room temperature.

| Device | $\Delta R_{NL}$ @ 20 K (Ω) | $\Delta R_{NL}$ @ RT (Ω) |
|---|---|---|
| A | 2.7 | 0.2 |
| B | 0.62 | — |
| C | 2.4 | — |
| D | — | 0.2 |
| E | 0.29 | 0.03 |
| F | 2.0 | — |
| G | 0.75 | — |
| H | 1.0 | 0.23 |
| I | 1.3 | 0.09 |

**Table S2.- Summary of spin-valve response**. Representative values of the spin valve signal, $\Delta R_{NL}$, observed for each device, for both 20 K and room temperature (RT). Cases where no measurement took place at a certain temperature are labelled '—'.

## Section 3. Contributions to the contact resistance $R_C$

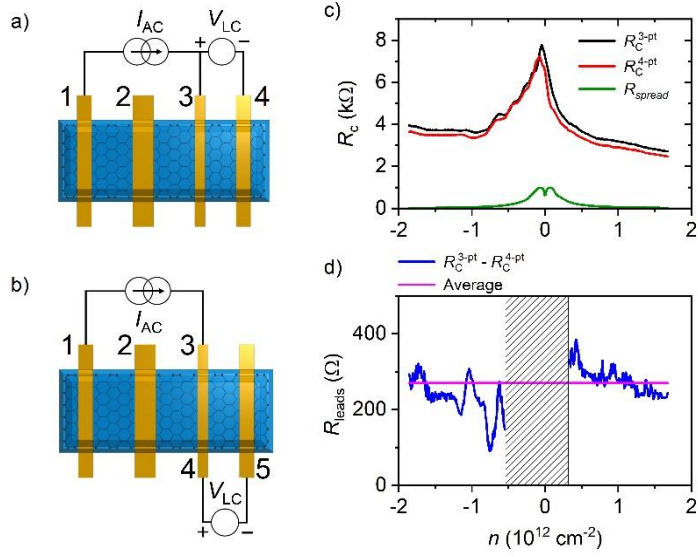

**Figure S3. Electrical characterisation of contact resistance. a, b,** Schematic of a 3-point (a) and 4-point (b) measurement configuration of $R_C$. **c,** Contact resistance as a function of carrier density for a typical contact, designed to perform measurements with both configurations (black and red curves respectively). Green curve is the calculated Maxwell spread resistance using the resistivity of the graphene channel. **d,** Estimation of the lead resistances by subtracting the 3-point and 4-point measurements.

We use both a 3-point method and a 4-point method, schematically represented in figure S3a and b respectively, to extract the contribution to the contact resistance coming from the leads connected to the contacts. Figure S3c shows the resistance of a typical contact measured with these two methods (black and red), as a function of carrier density. When both measurements are subtracted, we obtain a gate-independent value of around 250 Ω shown in figure S3d. We have found that this series resistance contribution is in the range of 100 to 300 Ω, for all our contacts. Therefore a value of 200 Ω has been deducted from the 3-point raw data for all devices. The shaded area in figure S3d indicates the range of carrier density where the deduction becomes unreliable due to variations in the junction resistance at low carrier density.

We analyse another possible contribution to the contact resistance, originating from charge current spreading, known as the Maxwell spreading resistance. We use the following approximation[5] for our device geometry,

$$R_{\text{spread}} = \frac{\rho}{2\pi} \ln\left(\frac{W}{l_{\text{m.f.p.}}}\right), \quad (1)$$

where $\rho$ is the resistivity of graphene, $W$ the width of the channel and $l_{\text{m.f.p.}}$ is the mean free path. The green curve in figure S3c shows the results of applying equation 1 to the measured resistivity as a function of carrier density. It is clear that its value is much lower than the contact resistance, demonstrating that the contribution of this term does not dominate the measured value of $R_C$.

## Section 4. Behaviour of one-dimensional magnetic contacts

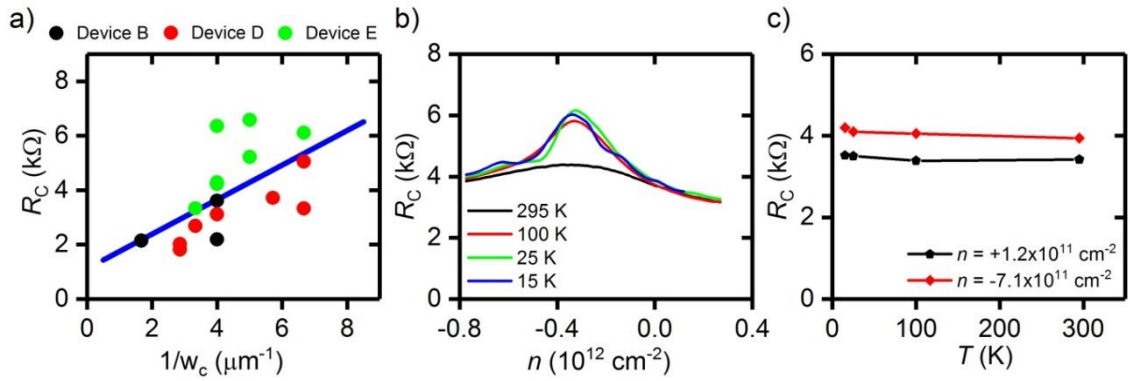

**Figure S4. 1D magnetic contacts. a,** Contact resistance scaling with the inverse of the contact width. $R_C$ is taken at 20 K and at moderate carrier density $n - n_0 \approx + 1.5x10^{12}$ cm$^{-2}$, where $n_0$ is the carrier density at the $R_C$ maxima. The blue line is a global linear fit of the data. **b,** Contact resistance as a function of carrier density, for different temperatures. **c,** Contact resistance as a function of temperature at two fixed carrier densities, for the same data as in **b**.

The width of the 1D contacts ($w_c$) was varied from 100 to 350 nm, to ensure different contacts reorient their magnetisation at different values of the field applied along their easy axis, $B_x$ (see Fig. 1(a)), for the purpose of spin valve measurements. This also enabled the possibility to characterise the scaling of their resistance with said widths. Figure S4a shows the scaling of contact resistance with the inverse of its width, measured on three different devices. The approximate scaling $R_c \propto 1/w_c$ is consistent with previous reports[8].

Figure S4b shows the dependence of the contact resistance with carrier density at different temperature, for a representative contact. The peak seen in all our contacts appears at $n \lesssim 0$, consistent with previous reports on the n-type doping of graphene in 2D Co contacts[9,10]. We have seen a significant increase in this peak resistance at low temperature, as depicted in Figure S4b. This is attributed to the emergence of ballistic transport at low temperature. At high carrier density, the contact resistance shows a negligible dependence on temperature (see Figure S4c), in agreement with the reported behaviour of 1D contacts fabricated with different materials[8]. Note the resistance of 1D contacts is material dependent. In the case of cobalt we found the contact resistance-width product to be in the range of 1 – 5 kΩ*μm.

## Section 5. Additional device with analysis of Hanle spin precession

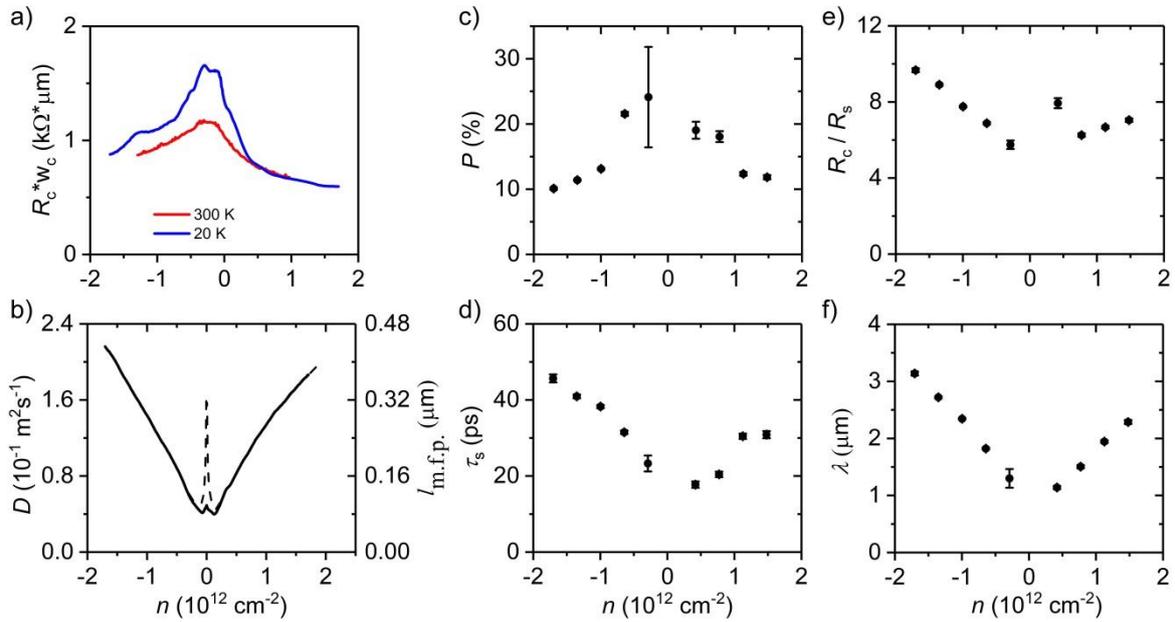

**Figure S5. Extraction of spin transport parameters from device C**. **a,** Contact resistance length product as a function of carrier density for a typical contact. **b,** Charge diffusion coefficient $D$ as a function of charge carrier density at 20 K. $D$ was calculated from the Einstein relation for an ideal DOS (dashed line) and for a DOS with a Gaussian broadening of ~40 meV (solid line). The mean free path (right axis) is extracted from the relation $l_{\text{m.f.p.}} = 2D/v_F$. **c, d, e, f,** Spin polarisation (**c**), Spin relaxation time (**d**), Contact resistance to spin resistance ratio (**e**), Spin relaxation length (**f**) as a function of carrier density at 20 K.

Figure S5 shows the spin transport parameters for device C, with a mobility of 30,000 cm$^2$V$^{-1}$s$^{-1}$ at 20 K. The spin parameters where extracted from the fitting of Hanle curves, as described in the main text. The range of values of the spin parameters $P$, $\tau_s$ and $\lambda$ and their electron-hole symmetry are commensurate with the devices presented in the main text, indicating consistent behaviour of the device architecture.

## Section 6. Exploring the mechanism of spin relaxation

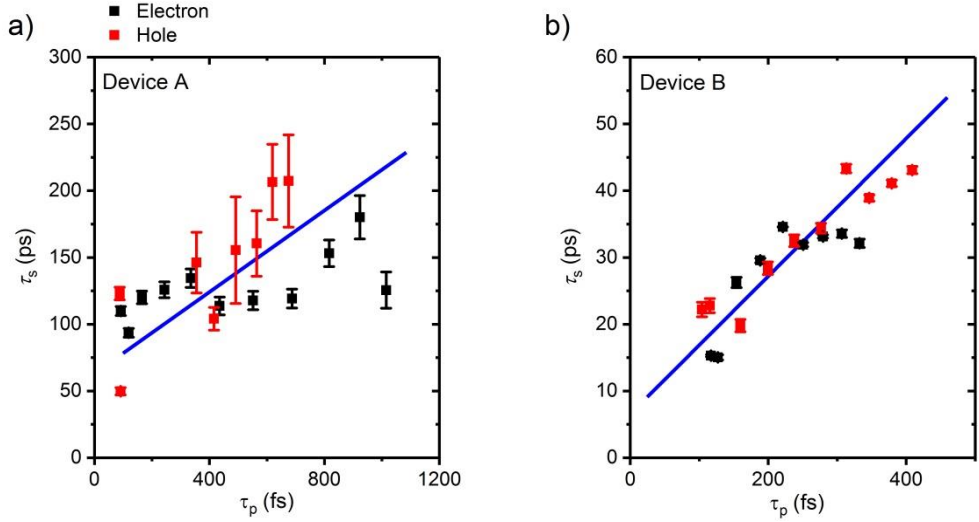

**Figure S6. Scaling of spin and momentum relaxation times**. **a, b,** A comparison between the spin- and momentum-relaxation times for devices A (**a**) and B (**b**). Black (red) squares are data from measurements made with the device in the electron (hole) regime. Blue lines are guides to the eye.

The momentum relaxation time is extracted from charge transport measurements, given by,

$$\tau_p = \frac{2D_C}{v_F^2}. \qquad (2)$$

Figure S6 shows the relation between the spin ($\tau_S$) and momentum ($\tau_p$) relaxation times for the devices A and B, discussed in the main text. This analysis shows larger momentum relaxation times are correlated with larger spin relaxation times, indicating an Elliot-Yafet[11,12] (EY)-like spin relaxation mechanism.

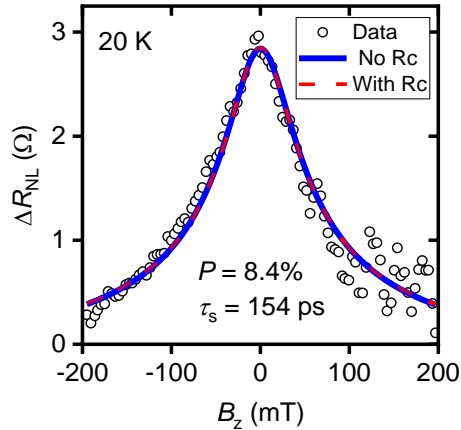

**Figure S7. Role of contact resistance**. Spin precession data, same as Fig. 3c, comparing a Hanle fit without considering contact resistance (blue line) and considering contact resistance (dashed red line).

The role of contact resistance on the extracted $\tau_S$ was assessed by performing fits to the Hanle data following the model by Fukuma et al.[13], which considers spin relaxation due to finite contact resistance (see Fig. S7). The fit including the contact resistance yields an extracted spin lifetime just 10 ps higher, which is well within the fitting error. Such a fit takes into account the role of relaxation via a finite contact resistance, thus giving access to a longer lifetime associated with relaxation only within the channel. Analysis for our dataset showed that the difference is consistently small, below the error, in agreement with the nature of our non-invasive contacts.

**Section 7. Long distance spin transport**

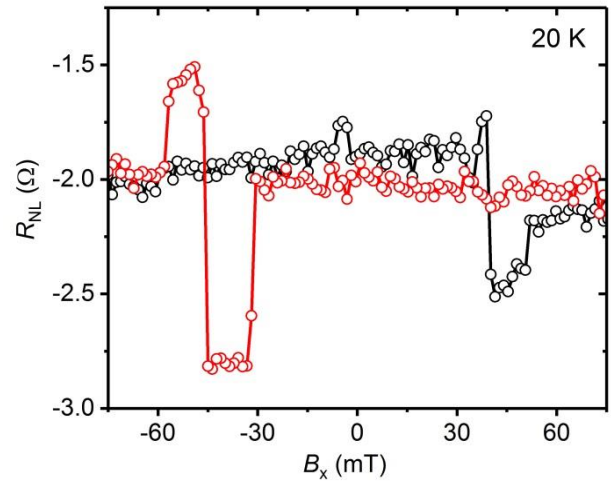

**Figure S8. Long distance spin valve**. Spin valve measurement at $n = -1.5 \times 10^{12}$ cm$^{-2}$ for a 12 µm long channel. The black (red) curve is for the up (down) sweep of the in-plane magnetic field.

Fig S8 shows the data from a spin valve measurement of a channel with a distance between injector and detector electrodes of $L = 12$ µm. Moreover, distributed along this length, there are five other 1D contacts across the channel. The observation of a spin-valve signal therefore confirms that our 1D magnetic contacts are of a non-invasive nature.

Furthermore, the observation of more than two switches (or more than two levels) in figure S8 indicates the detection of spin current by the closest reference contact to the detector, which is 3 µm further away. This implies spin transport for an even larger distance of 15 µm.

**References.**